\documentclass{PoS}

\usepackage{url}

\newcommand{\catnamelong}{Spitzer Data Fusion}
\newcommand{\catnameshort}{Spitzer Data Fusion}

\newcommand{\figdir}{./}
\newcommand{\myincludegraphics}[2]{\includegraphics*[#1]{#2}}

\title{The \catnamelong : Contents, Construction and Applications to Galaxy Evolution Studies}

\ShortTitle{The \catnamelong}

\author{\speaker{Mattia Vaccari}\thanks{HELP/SKA-SA Senior Research Fellow.}\\
           Department of Physics \& Astronomy, University of the Western Cape, Cape Town, South Africa\\
           INAF - Istituto di Radioastronomia, via Gobetti 101, 40129 Bologna, Italy\\
           E-mail: \email{mattia@mattiavaccari.net}}

\abstract{We present the \catnamelong, a database incorporating far-ultraviolet to far-infrared flux measurements as well as photometric and spectroscopic redshifts for 4.4 million IRAC-selected sources detected over 8 extragalactic fields covering 65 deg$^2$ observed by Spitzer in all IRAC and MIPS bands during its cryogenic mission. Deeper Spitzer observations carried out during its warm mission over 5 sub-fields as part of the SERVS project are also presented and analysed in a similar fashion, detecting 2.8 million IRAC-selected sources over 18 deg$^2$ and merging them with multi-wavelength catalogues within the SERVS Data Fusion. When combined with Herschel SPIRE surveys and radio continuum observations over the same fields, the \catnamelong\ and the SERVS Data Fusion provide an invaluable resource for multi-wavelength galaxy formation and evolution studies at infrared/millimetre/radio wavelengths. The catalogues and their future updates will be released at \url{http://www.mattiavaccari.net/df/} and on CDS/VizieR.}

\FullConference{EXTRA-RADSUR2015 (*)\\
		20--23 October 2015\\
		Bologna, Italy

                \bigskip
                \hrule
                \bigskip

                \textnormal{(*) This conference has been organized
                  with the support of the Ministry of Foreign Affairs
                  and International Cooperation, Directorate General
                  for the Country Promotion (Bilateral Grant Agreement
                  ZA14GR02 - Mapping the Universe on the Pathway to
                  SKA)}
                  }

\begin{document}



\maketitle

\section{Introduction}\label{intro.sec}
Notwithstanding the recent progress of astronomical archives, databases and virtual observatory tools, multi-wavelength catalogues readily useable for observational and interpretative work are difficult to come by for a variety of reasons, generally related with the different resolution and sensitivity achievable at different wavelengths and required to study galaxies at different redshifts. Here we present the first public release of the \catnamelong\ and of the SERVS Data Fusion, two Spitzer-selected multi-wavelength catalogues combining datasets extending from the far-ultraviolet to the far-infrared covering several of the most popular multi-wavelength extragalactic wide-area survey fields and thus providing a powerful resource for galaxy formation and evolution studies.
%
%
\section{Spitzer Data}\label{spitzer.sec}
In the course of its 2003-2009 cold mission the Spitzer mission \cite{Werner2004} has revolutionised galaxy formation and evolution studies, providing nearly optical-quality imaging over wide areas at mid-infrared wavelengths in its IRAC \cite{Fazio2004a} 3.6, 4.5, 5.8, 8.0 $\mu$m and MIPS \cite{Rieke2004} 24 $\mu$m channels as well as lower-resolution imaging in its MIPS 70 and 160 $\mu$m far-infrared channels. This wealth of mid-infrared and far-infrared wide-area imaging data, combined with  a number of imaging surveys at far-ultraviolet to near-infrared wavelengths, has allowed the astronomical community to greatly improve estimates of the stellar masses and star formation rates of galaxies up to high redshift. In producing the \catnameshort\ we attempted to provide an homogeneous database of far-ultraviolet-to-far-infrared photometry for Spitzer-selected sources covering several of the most popular wide-area extragalactic fields observed by Spitzer in its cryogenic mission.
\begin{figure*}
   \centering
   {\myincludegraphics{width=\textwidth}{\figdir{}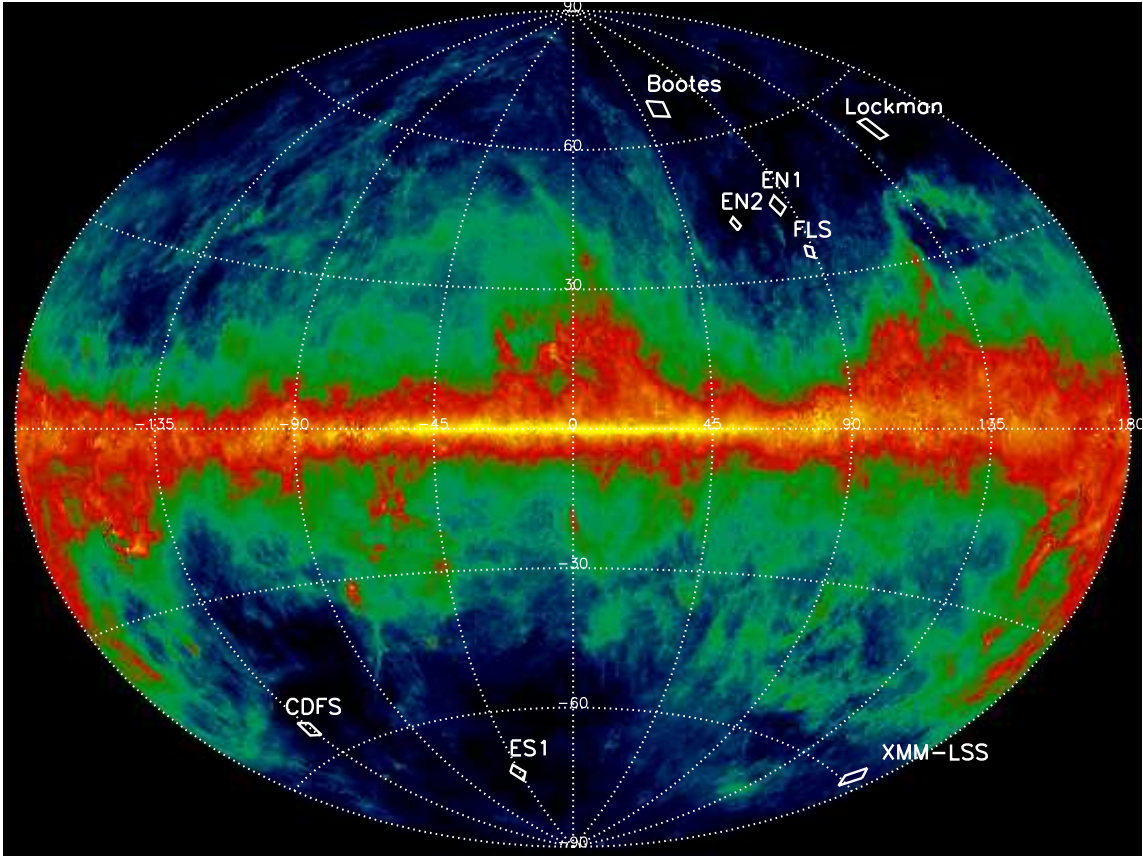}}
      \caption{\catnameshort\ fields overlaid on IRAS/COBE sky maps by \cite{Schlegel1998} in Galactic coordinates.}
   \label{spitzer-sky.fig}
   \end{figure*}
\subsection{SWIRE Fields}\label{swire.sec}
The Spitzer Wide-area Infrared Extragalactic (SWIRE, \cite{Lonsdale2003}) survey covers 50 deg$^2$ spread over six fields selected for their low levels of obscuration due to Galactic dust emission \cite{Schlegel1998}. As part of the \catnameshort, we undertook to improve upon available data products from the SWIRE survey and thus produce its Final Data Release. Here we thus briefly describe the properties of the SWIRE Final Data Release and how this was integrated within the \catnameshort. Further information about early SWIRE Data Processing is provided by \cite{Surace2005} and \cite{Shupe2008}.
%
%
\subsubsection{SWIRE IRAC and MIPS 24 $\mu$m Catalogues}\label{iracmips24.sec}
IRAC 4-band mosaics were produced using MOPEX \cite{MakovozMarleau2005} and IRAC sources were detected and their fluxes measured using SExtractor \cite{Bertin1996}. Kron fluxes (SExtractor's \texttt{AUTO} fluxes) as well as aperture fluxes were measured within five separate apertures, and aperture fluxes were corrected for PSF losses. Colour-Magnitude diagrams were constructed for various types of objects, in particular main-sequence stars. It was found that the scatter in these diagrams were minimised through the use of aper2, corresponding to a 1.9 arcsec aperture radius and roughly the size of the IRAC beam FWHM. We thus used IRAC aper2 fluxes as default IRAC fluxes to produce our database.

MIPS 24 $\mu$m mosaics were produced using MOPEX, as detailed by \cite{Shupe2008}. Note, however, that while \cite{Shupe2008} used SExtractor for MIPS 24 $\mu$m source extraction and flux measurement, the SWIRE Final Data Release adopted a more aggressive MIPS 24 $\mu$m source extraction technique than discussed by \cite{Shupe2008}, using APEX \cite{MakovozMarleau2005} rather than SExtractor. In the course of this work, however, we discovered (Shupe \& Vaccari, priv comm, see also \cite{Starikova2012}) how bright and/or extended MIPS 24 $\mu$m sources are often broken down in multiple components or lost altogether by the new APEX source extraction procedure. For the \catnameshort\ we thus devised an updated MIPS 24 $\mu$m source extraction procedure inspired by \cite{Shupe2008} which is both more reliable for bright and/or extended sources and as reliable as APEX for fainter and/or point-like sources. We thus re-extracted SWIRE MIPS 24 $\mu$m mosaics using SExtractor and produced Kron fluxes (SExtractor's \texttt{AUTO} fluxes) as well as aperture fluxes were measured within five separate apertures, and aperture fluxes were corrected for PSF losses. We then used the MIPS 24 $\mu$m aper2 (5.25" radius) aperture fluxes as default fluxes to produce our database. Measured fluxes were calibrated following the latest MIPS 24 $\mu$m nominal calibration \cite{Engelbracht2007} and colour-corrected so as to place them on a constant $\nu\,S_\nu$ scale which is appropriate for a wide range of Galaxy and AGN SEDs.
\subsubsection{SWIRE MIPS 70 and 160 $\mu$m Catalogues}\label{mipsge.sec}
MIPS 70 and 160 $\mu$m maps and catalogues were produced using MOPEX and APEX respectively. Total 70 and 160 $\mu$m fluxes were measured with APEX Point Response Function (PRF) fitting, and we used these as default fluxes to produce our database. Aperture fluxes were also measured within eight aperture radii, but they were not corrected for PSF losses. Photometric calibration was carried out following the latest MIPS instrumental calibration by \cite{Gordon2007} and \cite{Stansberry2007}, and fluxes were colour-corrected so as to put them (like their 24 $\mu$m counterparts) on a constant $\nu\,S_{\nu}$ scale. 
\subsection{Bootes Field}\label{bootes.sec}
The Bootes field was defined by the NOAO Deep Wide Field Survey (NDWFS) optical imaging covering $\sim$ 9 deg$^2$. The SDWFS (PI Daniel Stern, Spitzer PID 40839) and MAGES (PI Buell Jannuzi, Spitzer PID 50148) Spitzer programmes later provided IRAC and MIPS data respectively over the whole NDWFS field. In this work, SDWFS DR 1.1 IRAC public images by \cite{Ashby2009} were re-extracted using an updated IRAC SExtractor pipeline inspired by \cite{Surace2005} and the SWIRE Final Data Release, while MAGES MIPS data were re-reduced following the best practices employed in the reduction of similar data within the SWIRE fields at 24 $\mu$m (Shupe \& Vaccari, priv comm) as well as at 70 and 160 $\mu$m (Seymour \& Tugwell, priv comm). MIPS 24 $\mu$m sources were extracted using SExtractor and MIPS 70 and 160 $\mu$m sources were extracted using APEX, as done for the SWIRE fields. Flux calibration was also carried out as for the SWIRE fields.
\subsection{XFLS Field}\label{xfls.sec}
The Extragalactic First Look Survey (XFLS) field was defined by the Spitzer First Look Survey (FLS, PI Tom Soifer, Spitzer PID 26), which in its extragalactic survey component provided IRAC and MIPS data over $\sim$ 4 deg$^2$. In this work, as done for the Bootes field, XFLS IRAC public images by \cite{Lacy2005} were re-extracted using an updated IRAC SExtractor pipeline inspired by \cite{Surace2005} and by the SWIRE Final Data Release, while XFLS MIPS data were re-reduced following the best practices employed in the reduction of similar data within the SWIRE fields at 24 $\mu$m (Shupe \& Vaccari, priv comm) as well as at 70 and 160 $\mu$m (Seymour \& Tugwell, priv comm). MIPS 24 $\mu$m sources were extracted using SExtractor and MIPS 70 and 160 $\mu$m sources were extracted using APEX, as done for the SWIRE fields. Flux calibration was also carried out as for the SWIRE fields.
\subsection{Spitzer Sensitivity}
The sensitivity achieved in different Spitzer bands and fields was measured by placing a large number of random apertures across the images, measuring the fluxes within such apertures and fitting a Gaussian to the histograms of the measured fluxes. The dispersion (rms) of the resulting distribution, corrected for PSF losses, provides a reliable estimate of the image noise level. Tab.~\ref{spitzer-sens.tab} reports the $5\,\sigma$ flux limits resulting from this procedure.
\begin{table*}
\centering
\begin{small}
\begin{tabular}{|c|c|c|c|c|c|c|c|c|c|}
\hline
Channel & Unit & ES1 & XMM & CDFS & LH & EN1 & EN2 & Bootes & XFLS \\
\hline
IRAC1 & $\mu$Jy & 4.768 & 5.259 & 4.555 & 4.787 & 4.262 & 4.487 & 3.053 & 9.084 \\
IRAC2 & $\mu$Jy & 8.049 & 8.844 & 7.139 & 7.928 & 7.408 & 7.193 & 4.711 & 11.30 \\
IRAC3 & $\mu$Jy & 44.57 & 49.78 & 39.22 & 42.07 & 40.79 & 38.92 & 24.89 & 54.56 \\
IRAC4 & $\mu$Jy & 48.09 & 57.41 & 42.11 & 46.63 & 44.38 & 40.86 & 26.98 & 51.23 \\
MIPS1 & $\mu$Jy & 412.7 & 345.7 & 285.6 & 299.4 & 286.6 & 277.8 & 232.4 & 332.9 \\
MIPS2 &     mJy & 20.12 & 18.24 & 14.67 & 14.97 & 15.12 & 14.14 & 14.86 & 16.20 \\
MIPS3 &     mJy & 102.1 & 92.89 & 78.16 & 85.38 & 86.77 & 83.44 & 80.30 & 103.4 \\
\hline
\end{tabular}
\end{small}
\caption{Spitzer Data Fusion $5\,\sigma$ Flux limits in different bands and different fields. $\sigma$ was computed as described in the text.}
\label{spitzer-sens.tab}
\end{table*}
\section{Ancillary Datasets}\label{ancillary.sec}
A number of ancillary datasets, including photometry from the far-ultraviolet to the near-infrared and optical spectroscopy, were cross-correlated with the Spitzer catalogues described in the previous Section using a search radius of 1 arcsec. Tab.~\ref{spitzer-df.tab} summarises the ancillary datasets included in the \catnameshort\ for each field, including GALEX GR6Plus7 ultraviolet photometry \cite{Martin2005a}, SDSS DR12 optical photometry and spectroscopy \cite{Alam2015}, INT WFC optical photometry \cite{GonzalezSolares2011}, CFHTLS T0007 optical photometry, ESIS-WFI and VOICE-VST optical photometry, 2MASS PSC and XSC photometry \cite{Cutri2003}, UKIDSS DXS DR10Plus photometry \cite{Lawrence2007}, VIDEO VSA-DR4 photometry \cite{Jarvis2013}, IBIS DR2 photometry, HerMES PACS and SPIRE photometry \cite{Oliver2012}, the SWIRE photometric redshift catalogue \cite{RowanRobinson2013} and a collection of spectroscopic redshifts from NED as well as from the literature. The complete collection of spectroscopic redshifts is also made separately available at \url{http://www.mattiavaccari.net/df/specz/}.
\begin{table*}
\centering
\begin{small}
\begin{tabular}{|c|c|c|c|c|c|c|c|c|}
\hline
\textbf{Field} & \textbf{SDSS} & \textbf{INTWFC} & \textbf{CFHTLS} & \textbf{ESIS/VOICE} & \textbf{UKIDSS} & \textbf{VIDEO} &\textbf{IBIS} & \textbf{MRR} \\
\hline\hline
ES1     & -  & -  & -  & X & -  & X & - & X \\
XMM   & X & -  & X & -  & X & X & - & X \\
CDFS  & -  & -  & -  & X & -  & X & - & X \\
LH       & X & X & -  & -  & X & -  & - & X \\
EN1     & X & X & -  & -  & X & - & -  & X \\
EN1     & X & X & -  & -  & X & - & -  & X \\
Bootes & X & X & -  & -  & X & - & X & -  \\
XFLS   & X & X & -  & -  & X & - & -  & -  \\
\hline
\end{tabular}
\end{small}
\caption{Contents of the \catnameshort. All fields have GALEX, 2MASS, Spitzer IRAC and MIPS and Herschel PACS and SPIRE photometry as well as spectroscopic redshifts from multiple sources.}
\label{spitzer-df.tab}
\end{table*}
\section{Construction}\label{db-const.sec}
The \catnameshort\ is constructed by combining Spitzer IRAC and MIPS as well as ancillary catalogues according to the following positional association procedure.
\subsection{Spitzer Bandmerging}\label{spitzer-bandmerging.sec}
%
%
We combined the four IRAC single-band catalogues and the three MIPS single-band catalogues into a 7-band catalogue as follows. We first matched all IRAC 3.6 $\mu$m and IRAC 4.5 $\mu$m detections using a search radius of 1 arcsec, adopted a catalogue selection criterion keeping all sources detected at either 3.6 or 4.5 $\mu$m and defined the average of the 3.6 and 4.5 $\mu$m position as the source nominal coordinate. We then matched IRAC 5.8 $\mu$m and IRAC 8.0 $\mu$m sources against this source nominal coordinate using a 1.5 arcsec search radius. We then matched the MIPS 24 $\mu$m positions against the nominal coordinates of the IRAC 4-band catalogues using a search radius of 3 arcsec, and finally we matched the MIPS 70 $\mu$m and MIPS 160 $\mu$m coordinates against the MIPS 24 $\mu$m positions using a search radius of 6 and 12 arcsec respectively. We thus followed \cite{RowanRobinson2008} where a 24 $\mu$m detection is required to guarantee the reliability of 70 and 160 $\mu$m detections but also to more accurately pinpoint their positions. A cross-match reliability was defined for each MIPS band on the basis of the distance between the source positions and the source densities in the two bands to guide the user in selecting matched sources more or less aggressively. However, in all fields sources not detected at either 3.6 $\mu$m or 4.5 $\mu$m were discarded even when they might have had significant detections in other IRAC and/or MIPS channels. This was done to ensure, between other things, homogeneous sample selection and astrometric accuracy, and given the respective depths of the different images, only extreme examples of longer-wavelength sources with very peculiar colours may thus be missing an association at 3.6 $\mu$m and 4.5 $\mu$m. The full single-band catalogues at 24, 70, 160 $\mu$m are also separately provided to enable the investigation of individual sources: 24 $\mu$m catalogues for all fields and maps for Bootes and XFLS are available at \url{http://mattiavaccari.net/df/m24/}, while the 70 \& 160 $\mu$m catalogues for all fields and maps for Bootes and XFLS are available at \url{http://mattiavaccari.net/df/mipsge/}.
\subsection{Astrometric Registration}\label{db-astro.sec}
Both the Spitzer-selected catalogue and the ancillary catalogues are then astrometrically registered against 2MASS, which provides the best combination of astrometric accuracy and source density across the whole sky. 2MASS is also the astrometric reference frame adopted by the Spitzer Science Center for Spitzer operations and the Spitzer Heritage Archive, and its use should thus also minimise systematic astrometric offsets to be found between Spitzer-detected sources and their 2MASS $SNR>10$ counterparts. The median (RA,Dec) positional offsets against 2MASS are computed for high-significance ($SNR > 10$) detections and subtracted for each ancillary catalogue in each field, so that the resulting average offset of all bands against 2MASS is negligible. The positional normalised median absolute deviation (NMAD) for all sources after astrometric registration provides an indication of the level of astrometric accuracy of each ancillary catalogue at the faint end, and this is of order 0.3 and 0.9 arcsec for IRAC 3.6/4.5 $\mu$m and MIPS 24 $\mu$m respectively.
\subsection{Multi-Wavelength Bandmerging}\label{db-bandmerging.sec}
Following astrometric registration, ancillary catalogues are matched against Spitzer/IRAC positions using a simple nearest-neighbour association and an homogeneous search radius of 1 arcsec. While more refined multi-wavelength cross-correlation techniques following e.g. \cite{SutherlandSaunders1992} or \cite{Roseboom2009} may be useful in deeper Spitzer fields, the astrometric accuracies and source densities of individual catalogues under consideration are mostly well-matched to reliably provide a straightforward nearest-neighbour association. The number of matches obtained with each ancillary catalogue is reported in Tab.~\ref{db-counts.tab}. For each matched source in each ancillary catalogue the reliability of the match is defined based on the angular separation and on the source density of the ancillary catalogue.
%
%
\begin{table*}
\centering
\begin{tiny}
\begin{tabular}{cccccccccccc}
\hline
 Field &    IRAC &   MIPS &  MIPS &  MIPS &  GALEX &   SDSS &  INTWFC & 2MASS &  UKIDSS &  SpecZ &  Area \\
  Name & 3.6/4.5 $\mu$m & 24 $\mu$m & 70 $\mu$m & 160 $\mu$m & FUV/NUV & $ugriz$ & $ugriz$ & $J/H/K$ & $J/K$ & - & deg$^2$ \\
\hline
   ES1 &  391,518 &  39,491 &  1,653 &   695 &  43,308 &     NA &      NA & 10,716 &      NA &  10,092 &  7.00 \\
   XMM &  497,404 &  62,914 &  3,137 &  1,379 &  66,168 & 147,583 &      NA & 14,575 &  191689 &  42,919 &  9.35 \\
  CDFS &  464,084 &  67,584 &  3,418 &  1,431 &  59,976 &     NA &      NA & 12,785 &      NA &  28,166 &  8.15 \\
    LH &  660,053 &  91,779 &  4,722 &  2,076 & 82,819 & 205,283 &  407,862 & 16,919 &  390,552 &   6,905 & 11.50 \\
   EN1 &  573,843 &  85,394 &  4,041 &  1,865 &  66,867 & 199,067 &  344,817 & 20,959 &  432,835 &   4,048 &  9.65 \\
   EN2 &  273,650 &  40,867 &  1,929 &   769 &  34,470 &  97,956 &  163,151 & 11,263 &      NA &   1,276 &  4.40 \\
BOOTES & 1,301,829 & 117,083 &  4,609 &  2,921 & 100,918  & 227,132 &      NA &  7,320 &      NA &  20,808 & 11.10 \\
  XFLS &  228,354 &  32,225 &  2,050 &   282 &  23,164 &  73,316 &  116,725 & 11,018 &      NA &   2,724 &  4.00 \\
 Total & 4,390,735 & 537,337 & 25,559 & 11,418 & 477,690 & 950,337 & 1,032,555 & 40,019 & 1,015,076 & 116,938 & 65.15 \\ 
\hline
\end{tabular}
\end{tiny}
\caption{Number of sources contained in the \catnameshort\ at different wavelengths. The main catalogue selection (and thus the total number of sources) is at IRAC 3.6/4.5 $\mu$m, and the other catalogues are matched against IRAC 3.6/4.5 $\mu$m positions.}
\label{db-counts.tab}
\end{table*}

The \catnameshort\ aims to be the most extensive publicly available multi-wavelength catalogue of Spitzer-selected sources over the wide-area "Cosmic Windows" surveyed by Spitzer and now Herschel. This is by definition an ongoing work as we improve the data reduction of Spitzer data, including new data from Spitzer Warm surveys such as SERVS and DeepDrill and continue the collation of newly available multi-wavelength data. Here we describe the general properties of its first public release.

The \catnameshort\ DR1 will be made publicly available at \url{http://www.mattiavaccari.net/df/} as VO-compliant fits files which can be read in fortran with NASA's \texttt{cfitsio} library, in IDL with NASA's \texttt{astrolib} library as well as with \texttt{stilts/topcat} \cite{Taylor2005}, and it will be made available on CDS/Vizier (\url{http://vizier.u-strasbg.fr/viz-bin/VizieR}). For each field, the \catnameshort\ provides a "main" catalogue file and a number of "ancillary" catalogue files, one for each telescope/instrument used in ancillary observations. The "main" catalogue reports the position and one default flux measurement per each wavelength but only a limited number of other quantities and flags. Each "ancillary" catalogue contains several aperture and total flux measurements as well as a comprehensive set of photometric quantities and flags, allowing the expert user to make the most of ancillary observations.
\section{SERVS Catalogues and Data Fusion}\label{servs.sec}
The Spitzer Extragalactic Representative Volume Survey \cite{Mauduit2012} is the ideal successor of the SWIRE and S-COSMOS surveys in the Spitzer Warm era. Covering almost 40,\% of SWIRE footprint split over 5 fields at about the same depth as S-COSMOS, albeit only in the IRAC 3.6 and 4.5 micron channels, SERVS provides a mid-infrared counterpart to the UKIDSS-DXS and VISTA-VIDEO surveys, extending the redshift range where photometric redshifts and stellar masses can be reliably estimated and enabling the study of the evolution of massive galaxies at $1<z<5$. SERVS mosaics and highly reliable catalogues have been made publicly available and described as part of the SERVS1 DR1\footnote{\url{http://irsa.ipac.caltech.edu/data/SPITZER/SERVS/docs/SERVS_DR1_v1.4.pdf}}. To ensure the highest reliability and flux accuracy, DR1 catalogues were cut according to a number of criteria following the SNR and coverage levels. DR2 catalogues are effectively a more complete superset of DR1 catalogues obtained not applying either of these cuts while still retaining a $> 99\%$ reliability. The SERVS DR2 source extraction, catalogue production and bandmerging was therefore carried out as per the SERVS DR1 document. We used SExtractor \cite{Bertin1996} with a similar configuration to the one used for the \catnameshort\ to extract sources in the IRAC1 and IRAC2 images and measure five aperture fluxes corrected for PSF losses as well as Kron fluxes (SExtractor's \texttt{AUTO} fluxes). Aperture fluxes measured within 1.9 arcsec aperture radii (APER2) and Kron fluxes are recommended for studies of point-like and extended sources respectively. We then merged single-band detections into a two-band catalogue using a search radius of 1 arcsec. The SERVS DR2 single-band and two-band catalogues increase the number of IRAC-detected sources by a factor of 4 compared to SERVS DR1, and the SERVS DR2 Data Fusion is obtained matching SERVS DR2 catalogues with the ancillary catalogues making up the \catnameshort\ using a search radius of 1 arcsec. Tab.~4 describes the basic properties of the SERVS DR2 Catalogues, including $5\,\sigma$ sensitivities measured within 1.9 arcsec (APER2) aperture radii and corrected for PSF losses, whereas Tab.~\ref{spitzer-df.tab} described the contents of the SERVS DR2 Data Fusion. The SERVS Catalogues and the SERVS Data Fusion will be made publicly available at \url{http://www.mattiavaccari.net/df/} and on CDS/VizieR.
%
\begin{table*}
\centering
\begin{tabular}{|c|c|c|c|c|c|c|c|}
\hline
\textbf{Field} & \textbf{IRAC1AND2} & \textbf{IRAC1} & \textbf{IRAC2} & \textbf{IRAC1OR2} & $5\,\sigma_{IRAC1}$ & $5\,\sigma_{IRAC2}$ & \textbf{Area}\\
\hline\hline
ES1    & 301,861 & 450,197 & 457,596 & 605,932 & 2.20 & 2.53 & 3.0 \\
XMM  & 484,326 & 710,828 & 731,919 & 958,421 & 2.21 & 2.67 & 4.5 \\
CDFS & 425,970 & 620,841 & 634,320 & 829,191 & 2.12 & 2.48 & 4.5 \\
LH      & 466,433 & 684,599 & 732,936 & 951,102 & 2.21 & 2.70 & 4.0 \\
EN1    & 198,862 & 293,912 & 300,193 & 395,243 & 2.38 & 2.57 & 2.0 \\
Total & 1,877,452 & 2,760,377 & 2,856,964 & 3,739,889 & - & - &18.0 \\
\hline
\end{tabular}
\label{servs-cats.tab}
\caption{Basic Properties of SERVS DR2 Catalogues. Number of sources detected in both, individual and either the IRAC1 and IRAC2 bands, Flux Accuracy and Sky Coverage in different fields.}
\end{table*}
\section{Applications}\label{applications.sec}
The \catnameshort\ and the SERVS Data Fusion were developed within the SWIRE, SERVS and HerMES collaborations to meet the need for an homogeneous multi-wavelength far-ultraviolet-to-far-infrared catalogue of Spitzer-selected sources. It thus fills a gap preventing the effective exploitation of Spitzer and Herschel cosmological surveys, and in so doing has produced a number of useful applications.

Of particular importance are the applications to the Herschel Multi-Tiered Extragalactic Survey (HerMES, \cite{Oliver2012}), where the \catnameshort\ was extensively since the Herschel Science Demonstration Phase through mission operations and into the post-mission phase to best exploit SPIRE sub-millimetre observations. Applications ranged from the optimal astrometric registration of Herschel map-making by \cite{Levenson2010} to the optimisation of Herschel source extraction techniques based on shorter-wavelength positional priors by \cite{Roseboom2010,Roseboom2012}, from
the determination of the local star formation rate density \cite{Marchetti2016} to its evolution up to high redshift 
\cite{Eales2010b,RowanRobinson2016}, from studies of the evolution of dust temperature with redshift \cite{Hwang2010,Symeonidis2013} to linking the X-ray and far-infrared properties of star-forming galaxies \cite{Symeonidis2011,Symeonidis2014}.

A further range of applications include studying the link between far-infrared and radio continuum emission \cite{Ivison2010,Jarvis2010,Smith2014} and investigating the nature of radio continuum sources and their evolution through multi-wavelength photometry \cite{Smolcic2009a,Smolcic2009b}. \cite{Luchsinger2015,Rawlings2015,Whittam2015} have e.g. recently combined the \catnameshort\ and the SERVS Data Fusion with multi-wavelength radio continuum imaging to investigate the fraction of radio sources associated with Active Galactic Nuclei and Star Forming Galaxies. Much improved studies of the Far-Infrared/Radio Correlation as well as of the Cosmic Star Formation and Black Hole Accretion History of the Universe will be enabled by the upcoming deep and wide surveys planned with the SKA pathfinders and precursors \cite{P&S2015,Jarvis2015,McAlpine2015,Smolcic2015,Zwart2015}. The \catnameshort\ and the SERVS Data Fusion and their future incarnations will provide a precious resource for their timely exploitation.
\section{Conclusions}\label{conclusions.sec}
We presented the \catnamelong, a database incorporating far-ultraviolet to far-infrared flux measurements as well as photometric and spectroscopic redshifts for 4.4 million IRAC-selected sources detected over 8 extragalactic fields covering 65 deg$^2$ observed by Spitzer in all seven IRAC and MIPS during its cryogenic mission. Deeper Spitzer observations carried out during its warm mission over 5 sub-fields as part of the SERVS project are also presented and analysed in a similar fashion, detecting 2.8 million IRAC-selected sources over 18 deg$^2$ and merging them with multi-wavelength catalogues within the SERVS Data Fusion. When combined with Herschel SPIRE surveys and radio continuum observations over the same fields, the \catnamelong\ and the SERVS Data Fusion provide an invaluable resource for multi-wavelength galaxy formation and evolution studies at infrared/millimetre/radio wavelengths. The catalogues and their future updates will be released at \url{http://www.mattiavaccari.net/df/} and on CDS/VizieR. They will be further expanded and refined as part of the Herschel Extragalactic Legacy Project (HELP, \cite{Vaccari2016}), an EC-REA FP7-SPACE project bringing together multi-wavelength surveys in Herschel extragalactic fields.
\section*{Acknowledgements}
We gratefully acknowledge the continuing work of the Spitzer team as well as of the many teams involved in delivering the ancillary datasets used in this work. While a number of software tools were used in producing the \catnameshort, \texttt{stilts/topcat} \cite{Taylor2005} was probably the only one which (new graphics notwithstanding:-) was more often than not a pleasure to use.

Mattia Vaccari is supported by the European Commission's REA (FP7-SPACE-2013-1 GA 607254 - Herschel Extragalactic Legacy Project), South Africa's DST (DST/CON 0134/2014) and Italy's MAECI (PGR GA ZA14GR02 - Mapping the Universe on the Pathway to SKA).
\section*{Appendix : Spitzer Data Fusion Database of Instrumental Transmissions}\label{filters.sec}
The \catnameshort\ merges observations obtained with a wide variety of ground-based and space-based telescopes, and the SED template fitting process required to make sense of the physical properties of detected sources therefore combines photometric observations obtained from most of these instruments. It is thus useful to assemble accurate Instrumental Transmissions (incorporating the Telescope, Filter and Detector transmissions) for the various datasets. The filter database originally distributed with the \texttt{Hyperz} package and using the same filter specification format as the \texttt{LePhare} package was extensively updated with old and new filters gathered from the literature and online to provide a reliable instrumental transmission. The latest public release of the database can be obtained at \url{http://www.mattiavaccari.net/df/filters/}. A look-up table indicating the filters corresponding to the data contained within the \catnameshort\ and SERVS Data Fusion will also be made available upon its public release.

\end{document}